\documentclass[ALICE,manyauthors]{cernphprep}
\usepackage[comma,square,numbers,sort&compress]{natbib}
\usepackage{graphicx}
\usepackage{lineno}
\usepackage{xspace}
\usepackage{hyperref}
\usepackage{siunitx}
\DeclareSIUnit\clight{\text{\ensuremath{c}}}
\DeclareSIUnit\eVc{\eV/\clight} 
\DeclareSIUnit\keVc{\keV/\clight} 
\DeclareSIUnit\MeVc{\MeV/\clight} 
\DeclareSIUnit\MeVcc{\MeV/\clight\squared} 
\DeclareSIUnit\GeVc{\GeV/\clight} 
\DeclareSIUnit\GeVcc{\GeV/\clight\squared} 
\usepackage[Symbol]{upgreek}
\usepackage[T1]{fontenc}

\begin{document}
%

\newcommand{\pp}           {pp\xspace}

\newcommand{\roots}        {\ensuremath{\sqrt{s}}\xspace}
\newcommand{\pt}           {\ensuremath{p_\mathrm{ T}}\xspace}
\newcommand{\pT}{\ensuremath{p_{\mathrm{T}}}\xspace} 
\newcommand{\px}           {\ensuremath{p_\mathrm{ x}}\xspace}
\newcommand{\py}           {\ensuremath{p_\mathrm{ y}}\xspace}
\newcommand{\etarange}[1]  {\mbox{$\left| \eta \right| < #1$}}
\newcommand{\yrange}[1]    {\mbox{$\left| y \right| < #1$}}
\newcommand{\dndy}         {\ensuremath{\mathrm{d}N_\mathrm{ch}/\mathrm{d}y}\xspace}
\newcommand{\dndeta}       {\ensuremath{\mathrm{d}N_\mathrm{ch}/\mathrm{d}\eta}\xspace}
\newcommand{\avdndeta}     {\ensuremath{\langle\dndeta\rangle}\xspace}
\newcommand{\dNdy}         {\ensuremath{\mathrm{d}N_\mathrm{ch}/\mathrm{d}y}\xspace}
\newcommand{\dEdx}         {\ensuremath{\textrm{d}E/\textrm{d}x}\xspace}
\newcommand{\kstar}        {\ensuremath{k^*}\xspace}
\newcommand{\rstar}        {\ensuremath{r^*}\xspace}
\newcommand{\mt}           {\ensuremath{m_{\mathrm{T}}}\xspace}
\newcommand{\St}           {\ensuremath{S_{\mathrm{T}}}\xspace}
\newcommand{\Np}           {\ensuremath{N_{\mathrm{prompt}}}\xspace}
\newcommand{\Nnp}           {\ensuremath{N_{\mathrm{non\text{-}prompt}}}\xspace}
\newcommand{\rawY}[1]          {\ensuremath{Y_{#1}}\xspace}
\newcommand{\effNP}[1]         {\ensuremath{(\mathrm{Acc}\times\epsilon)^\mathrm{non\text{-}prompt}_{#1}}\xspace}
\newcommand{\effP}[1]          {\ensuremath{(\mathrm{Acc}\times\epsilon)^\mathrm{prompt}_{#1}}\xspace}
\newcommand{\AccEff}[1]          {\ensuremath{(\mathrm{Acc}\times\epsilon)_{#1}}\xspace}
\newcommand{\fnonprompt}   {\ensuremath{f_\mathrm{non\text{-}prompt}}\xspace}

\newcommand{\tev}          {\ensuremath{\mathrm{TeV}}\xspace}
\newcommand{\gev}          {\ensuremath{\mathrm{GeV}}\xspace}
\newcommand{\mev}          {\ensuremath{\mathrm{MeV}}\xspace}
\newcommand{\lumi}         {\ensuremath{\mathcal{L}_\mathrm{int}}\xspace}
\newcommand{\mum}          {\ensuremath{\mathrm{\upmu m}}\xspace}
\newcommand{\centm}        {\ensuremath{\mathrm{cm}}\xspace}
\newcommand{\invnb}        {\ensuremath{\mathrm{nb}^{-1}}\xspace}
\newcommand{\fm}        {\ensuremath{\mathrm{fm}}\xspace}

\newcommand{\ITS}          {\ensuremath{\mathrm{ITS}}\xspace}
\newcommand{\TOF}          {\ensuremath{\mathrm{TOF}}\xspace}
\newcommand{\ZDC}          {\ensuremath{\mathrm{ZDC}}\xspace}
\newcommand{\ZDCs}         {\ensuremath{\mathrm{ZDCs}}\xspace}
\newcommand{\ZNA}          {\ensuremath{\mathrm{ZNA}}\xspace}
\newcommand{\ZNC}          {\ensuremath{\mathrm{ZNC}}\xspace}
\newcommand{\SPD}          {\ensuremath{\mathrm{SPD}}\xspace}
\newcommand{\SDD}          {\ensuremath{\mathrm{SDD}}\xspace}
\newcommand{\SSD}          {\ensuremath{\mathrm{SSD}}\xspace}
\newcommand{\TPC}          {\ensuremath{\mathrm{TPC}}\xspace}
\newcommand{\TRD}          {\ensuremath{\mathrm{TRD}}\xspace}
\newcommand{\VZERO}        {\ensuremath{\mathrm{V0}}\xspace}
\newcommand{\VZEROA}       {\ensuremath{\mathrm{V0A}}\xspace}
\newcommand{\VZEROC}       {\ensuremath{\mathrm{V0C}}\xspace}
\newcommand{\Vdecay} 	   {\ensuremath{V^{0}}\xspace}

\newcommand{\ee}           {\ensuremath{\mathrm{e}^{+}\mathrm{e}^{-}}} 
\newcommand{\pip}          {\ensuremath{\uppi^{+}}\xspace}
\newcommand{\pim}          {\ensuremath{\uppi^{-}}\xspace}
\newcommand{\kap}          {\ensuremath{\mathrm{K}^{+}}\xspace}
\newcommand{\kam}          {\ensuremath{\mathrm{K}^{-}}\xspace}
\newcommand{\KK}           {\ensuremath{\mathrm{K}^{+}\mathrm{K}^{-}}\xspace}
\newcommand{\KKbar}           {\ensuremath{\mathrm{K}\overline{\mathrm{K}}}\xspace}
\newcommand{\phiPart}      {\ensuremath{\upphi}\xspace}
\newcommand{\pbar}         {\ensuremath{\mathrm\overline{p}}\xspace}
\newcommand{\kzero}        {\ensuremath{{\mathrm K}^{0}_{\mathrm{S}}}\xspace}
\newcommand{\lmb}          {\ensuremath{\Lambda}\xspace}
\newcommand{\almb}         {\ensuremath{\overline{\Lambda}}\xspace}
\newcommand{\Om}           {\ensuremath{\Omega^-}\xspace}
\newcommand{\Mo}           {\ensuremath{\overline{\Omega}^+}\xspace}
\newcommand{\X}            {\ensuremath{\Xi^-}\xspace}
\newcommand{\Ix}           {\ensuremath{\overline{\Xi}^+}\xspace}
\newcommand{\Xis}          {\ensuremath{\Xi^{\pm}}\xspace}
\newcommand{\Oms}          {\ensuremath{\Omega^{\pm}}\xspace}

\newcommand{\NDbar}{\ensuremath{\mbox{N--}\overline{\mathrm{D}}}\xspace}

\newcommand{\DplustoKpipi}{\ensuremath{\mathrm{D^\pm\to K^\mp\uppi^\pm\uppi^\pm}}\xspace}
\newcommand{\DstartoDpluspi}{\ensuremath{\mathrm{D^{*\pm}\to D^\pm\uppi^0}}\xspace}
\newcommand{\DstartoDplusGamma}{\ensuremath{\mathrm{D^{*\pm}\to D^\pm\upgamma}}\xspace}
\newcommand{\DstarminustoDzeropi}{\ensuremath{\mathrm{D^{*-}\to \overline{D}^0\uppi^-\to K^+\uppi^-\uppi^-}}\xspace}
\newcommand{\DstartoDplus}{\ensuremath{\mathrm{D^{*\pm}\to D^\pm X}}\xspace}
\newcommand{\DstartoDminus}{\ensuremath{\mathrm{D^{*-}\to D^- X}}\xspace}
\newcommand{\Dminus}{\ensuremath{\mathrm{D}^{-}}\xspace}
\newcommand{\Dplus}{\ensuremath{\mathrm{D}^{+}}\xspace}
\newcommand{\Dplusminus}{\ensuremath{\mathrm{D}^\pm}\xspace}
\newcommand{\Dstarplus}{\ensuremath{\mathrm{D}^{*+}}\xspace}
\newcommand{\Dstarplusminus}{\ensuremath{\mathrm{D}^{*\pm}}\xspace}
\newcommand{\Dstarminus}{\ensuremath{\mathrm{D}^{*-}}\xspace}
\newcommand{\Dstar}{\ensuremath{\mathrm{D}^*}\xspace}
\newcommand{\Dzero}{\ensuremath{\mathrm{D}^0}\xspace}
\newcommand{\Dzerobar}{\ensuremath{\mathrm{\overline{D}}^0}\xspace}

\newcommand{\rCore} {\ensuremath{R_{\mathrm{core}} = (0.80\pm0.08)}\,fm\xspace}
\newcommand{\rEff} {\ensuremath{R_{\mathrm{eff}} = (0.89\pm0.08)}\,fm\xspace}

\newcommand{\pP}{\ensuremath{\mathrm{pp}}\xspace}
\newcommand{\ApAp}{\ensuremath{\mbox{\overline{p}\overline{p}}}\xspace}
\newcommand{\nDbar} {\ensuremath{\mbox{n}\overline{\mathrm{D}}^{0}}\xspace}
\newcommand{\pD} {\ensuremath{\mathrm{pD^-}}\xspace}
\newcommand{\DN} {\ensuremath{\mathrm{ND}}\xspace}
\newcommand{\DbarN} {\ensuremath{\mathrm{N\overline{D}}}\xspace}
\newcommand{\apaD} {\ensuremath{\mathrm{\overline{p}D}^+}\xspace}
\newcommand{\pDcomb} {\ensuremath{\mathrm{pD}^- \oplus \mathrm{\overline{p}D}^+}\xspace}

\newcommand{\pKpipi} {\ensuremath{\mathrm{p(K}^{+}\uppi^{-}\uppi^{-}\mathrm{)}}\xspace}
\newcommand{\Kpipi} {\ensuremath{\mbox{(K}^{+}\uppi^{-}\uppi^{-}\mathrm{)}}\xspace}

\newcommand{\pDstar} {\ensuremath{\mathrm{pD^{*-}}}\xspace}

\newcommand{\ccbar}           {\ensuremath{\mathrm{c\overline{c}}}\xspace} 
\newcommand{\bbbar}           {\ensuremath{\mathrm{b\overline{b}}}\xspace} 

\newcommand{\Pc}[1]           {\ensuremath{\mathrm{P_\mathrm{c}(#1)}}\xspace} 

\newcommand{\NsigmaCoulomb}         {\ensuremath{(1.1\text{--}1.5)}\xspace}
\newcommand{\NsigmaHaidenbauer}     {\ensuremath{(1.2\text{--}1.5)}\xspace}
\newcommand{\NsigmaHaidenbauerMod}  {\ensuremath{(0.8\text{--}1.3)}\xspace}
\newcommand{\NsigmaHofmannLutz}     {\ensuremath{(1.3\text{--}1.6)}\xspace}
\newcommand{\NsigmaYamaguchi}       {\ensuremath{(0.6\text{--}1.1)}\xspace}
\newcommand{\NsigmaFontouraOne}        {\ensuremath{(1.0\text{--}1.5)}\xspace}
\newcommand{\NsigmaFontouraTwo}        {\ensuremath{(1.1\text{--}1.5)}\xspace}

\newcommand{\resultPot}             {\ensuremath{V_{\mathrm{I}=0} \in [-1450,-1050]~\si{\MeV}}\xspace}
\newcommand{\resultInvScatt}        {\ensuremath{f_{0,~\mathrm{I}=0}^{-1} \in [-0.4,0.9]~\si{fm^{-1}}}\xspace}
\newcommand{\scattLen}              {\ensuremath{f_{0,~\mathrm{I}=0}^{-1}}\xspace}
\newcommand{\reff}              {\ensuremath{R_\mathrm{eff}}\xspace}
\newcommand{\etal}              {et al\xspace}

\begin{titlepage}
\PHyear{2022}       
\PHnumber{006}      
\PHdate{12 January}  

\title{First study of the two-body scattering involving charm hadrons}
\ShortTitle{First study of the two-body scattering involving charm hadrons}   

\Collaboration{ALICE Collaboration\thanks{See Appendix~\ref{app:collab} for the list of collaboration members}}
\ShortAuthor{ALICE Collaboration} 

\begin{abstract}
This article presents the first measurement of the interaction between charm hadrons and nucleons.
The two-particle momentum correlations of \pD and \apaD pairs are measured by the ALICE Collaboration in high-multiplicity
pp collisions at $\roots = 13~\tev$. The data are compatible with the Coulomb-only interaction hypothesis within $\NsigmaCoulomb\sigma$.
The level of agreement slightly 
improves if an attractive nucleon(N)$\overline{\mathrm{D}}$ strong interaction is 
considered, in contrast to most model predictions which suggest an overall repulsive interaction. This measurement allows for the first time an estimation of the $68$\% confidence level interval for the isospin $\mathrm{I}=0$ inverse scattering length of the \DbarN state \resultInvScatt, assuming negligible interaction for the isospin $\mathrm{I}=1$ channel. 
\end{abstract}
\end{titlepage}

\setcounter{page}{2} 


\section{Introduction}
The study of the residual strong interaction among hadrons is a very active field within nuclear 
physics. 
This interaction can lead to the formation of bound states, such as nuclei, or molecular states as, for example, the $\Lambda(1405)$, which is considered as being generated from the attractive forces in the nucleon(N)$\overline{\mathrm{K}}$--$\Sigma\uppi$ channels~\cite{Hyodo:2011ur,Meissner:2020khl,Mai:2020ltx,Hyodo:2020czb}. 
One of the most fervent discussions in this context is nowadays revolving around systems involving charm mesons (D, \Dstar). 
Studies of their interaction are motivated by the observation of several new states with hidden charm and/or beauty (so-called XYZ states)~\cite{Swanson:2006st,Guo:2017jvc,Hosaka:2016pey,Brambilla:2019esw,Aaij:2019evc}, as well as with open charm such as the $\mathrm{T_{cc}}^+$~\cite{LHCb:2021vvq,LHCb:2021auc}, and also of  pentaquark states like \Pc{4380} and \Pc{4450}~\cite{Aaij:2015tga,LHCb:2019kea}. These exotic hadrons can be described as compact multiquark states in the context of the constituent-quark model~\cite{GellMann:1964nj}, but are also considered as natural candidates for loosely bound molecular states~\cite{Swanson:2006st,Guo:2017jvc}. For example, the structure of the $\chi_\mathrm{c1}$(3872) (formerly X(3872)) has been interpreted as a $\overline{\mathrm{D}}\mathrm{D}^*$/D$\overline{\mathrm{D}}{}^*$ molecular state or as a tetraquark~\cite{Chen:2016qju}. Currently, definite conclusions are difficult to draw because of the lack of any direct experimental information on the D$\overline{\mathrm{D}}{}^*$ strong interaction. 
Strong support for the molecular nature of the $\Lambda(1405)$ came not least from low-energy N$\overline{\mathrm{K}}$ scattering data and information on the p$\overline{\mathrm{K}}$ scattering length from kaonic hydrogen atoms~\cite{Borasoy:2004kk,SIDDHARTA:2011dsy,Ikeda:2011pi,Ikeda:2012au}. Hence, a determination of the scattering parameters of systems involving D and/or \Dstar mesons are pivotal to advance in the interpretation of the many observed states. The first step in this direction is the investigation of the interaction between the p(uud)$\Dminus(\mathrm{\overline{c}d})$ pair and its charge conjugate. This interaction does not couple to the lower energy meson--baryon channels since no $\mathrm{q\overline{q}}$ annihilation can occur. A measurement of this interaction is also an essential reference for the study of the in-medium D- and \Dstar-meson properties~\cite{He:2011yi}. Similarly to kaons and antikaons, it is theoretically predicted that possible modifications of the charm-meson spectral function at large baryonic densities can be connected to a decrease of the chiral condensate, thus providing sensitivity to chiral-symmetry restoration~\cite{Tolos:2009nn}.

So far, the topic of the strong interaction between hadrons containing charm quarks was addressed only from a theoretical point of view~\cite{Haidenbauer,Hofmann:2005sw,Fontoura,Yamaguchi} by employing different effective models anchored to the successful description of other baryon--meson final states, such as the N$\overline{\mathrm{K}}$ and NK systems, while data are missing. Scattering experiments~\cite{Stoks1993PartialwaveAO} 
and systematic studies of stable and unstable nuclei~\cite{Nowacki:2021fjw}, accompanied by sophisticated calculations achieved within effective field theories~\cite{Hebeler:2015hla,Gebrerufael:2016xih}, allowed us to reach a solid comprehension of the interaction 
among nucleons. When extending these studies to interactions including strange hadrons, the average properties of the interactions of some strange nucleon--hadron combinations (p$\mathrm{K^\pm}$~\cite{Abrams:1965zz,Hyslop:1992cs,Olin:2001ts}, p$\Lambda$, and p$\Sigma^0$~\cite{Eisele:1971mk,Alexander:1969cx,SechiZorn:1969hk}) could be gauged with the help of scattering data and measurements of kaonic atoms~\cite{Hrtankova:2017zxw}.
The study of $\Lambda$ hypernuclei~\cite{Feliciello:2015dua} led to the extraction of an average attractive potential.
The situation has drastically changed in recent years, thanks to the novel employment of the femtoscopy technique~\cite{Pratt:1986cc} in pp and p--Pb collisions at the LHC applied to almost all combinations of protons and strange hadrons~\cite{ALICE:2018ysd}. The ALICE Collaboration could precisely study the following interactions: pp, pK$^{\pm}$, p$\Lambda$, p$\overline{\Lambda}$, p$\Sigma^0$, $\Lambda\Lambda$, $\Lambda\overline{\Lambda}$,  p$\Xi^-$, p$\Omega^-$, and p$\upphi$ ~\cite{ALICE:2018ysd, FemtoKaon,FemtopLambda,FemtopSigma,FemtoLambdaLambda,FemtopXi,FemtoNature,FemtopPhi, FemtoAnti}. 
Since conventional scattering experiments cannot be performed with D mesons and charm nuclei~\cite{Hosaka:2016ypm} have not been discovered yet (searches for charm nuclear states are included in the scientific program of the Japan Proton Accelerator Research Complex~\cite{Noumi:2017hbc}), the femtoscopy technique can be employed to study the ND and \DbarN interactions. In this article, the first measurement of the strong interaction between a \Dminus meson and a proton is reported. This pioneering analysis employs \Dminus instead of the more abundantly produced \Dzerobar mesons because of the smaller contribution from decays of excited charm states and the possibility to separate particles and antiparticles without ambiguity.

\section{Experimental apparatus and data samples}
The analysis was performed using a sample of high-multiplicity pp collisions at $\roots = 13~\tev$ collected by ALICE~\cite{ALICE,ALICE:2014sbx} during the LHC Run 2 (2016--2018). The main detectors used for this analysis to reconstruct and identify the protons and the D-meson decay products are the Inner Tracking System (ITS)~\cite{ALICE:2010tia}, the Time Projection Chamber (TPC)~\cite{Alme:2010ke} and the Time-Of-Flight (TOF) detector~\cite{Akindinov:2013tea}. They are located inside a large solenoidal magnet providing a uniform magnetic field of $0.5$~T parallel to the LHC beam direction and cover the pseudorapidity interval $|\eta|<0.9$.
The events were recorded with a high-multiplicity trigger relying on the measured signal amplitudes in the V0 detector, which consists of two scintillator arrays covering the pseudorapidity intervals $-3.7 < \eta < -1.7$ and $2.8 < \eta < 5.1$~\cite{ALICE:2013axi}. The collected data sample corresponds to the 0.17\% highest-multiplicity events out of all inelastic collisions with at least one charged particle in the pseudorapidity range $|\eta|<1$ (denoted as INEL $>$ 0).
Events were further selected offline in order to remove machine-induced backgrounds~\cite{ALICE:2014sbx}. The events were required to have a reconstructed collision vertex located within $\pm10~\centm$ from the center of the detector along the beam-line direction to maintain a uniform acceptance. Events with multiple primary vertices (pileup), reconstructed from track segments measured with the two innermost ITS layers, were rejected. The remaining undetected pileup is of the order of $1\%$ and therefore negligible in the analysis. After these selections, the analyzed data sample consists of about $10^9$ events.
The Monte Carlo (MC) samples used in this analysis consist of pp collisions simulated using the \textsc{Pythia~8.243} event generator~\cite{Sjostrand:2006za, Sjostrand:2014zea} with the Monash-13 tune~\cite{Skands:2014pea} and \textsc{Geant3}~\cite{GEANT} for the propagation of the generated particles through the detector.

\section{Data analysis}
\subsection{Selection of proton and $\pmb{\Dplusminus}$-meson candidates}
\label{sec:selection}
The proton candidates are selected according to the methods described in~\cite{ALICE:2018ysd}.
Charged-particle tracks reconstructed with the \TPC are required to have transverse momentum $0.5<\pT<4.05~\gev/c$ and pseudorapidity \etarange{0.8}.
Particle identification (PID) is conducted by measuring the specific energy loss and the time of flight with the \TPC and \TOF detectors, respectively. The selection is based on the deviation $n_\sigma$ between the measured and expected values for protons, normalized by the detector resolution $\sigma$. For proton candidates with a momentum $p < \SI{0.75}{\gev/c}$, only the \TPC is used by requiring $|n_\sigma^\TPC| < 3$, while for larger momenta the PID information of \TPC and \TOF are combined and tracks are accepted only if the condition $\sqrt{(n_\sigma^\TPC)^2+(n_\sigma^\TOF)^2} < 3$ is fulfilled. With these selection criteria, the purity of the proton sample averaged over $\pt$ is $P_\mathrm{p} = 98\%$~\cite{ALICE:2018ysd}. The contribution of secondary protons originating from weak decays or interactions with the detector material is assessed by using MC template fits to the measured distribution of the distance of closest approach of the track to the primary vertex. The estimated average fraction of primary protons is 86\%~\cite{ALICE:2018ysd}.

The \Dplusminus mesons are reconstructed via their hadronic decay channel \DplustoKpipi, having a branching ratio $\mathrm{BR}=(9.38\pm0.15)\%$~\cite{Zyla:2020zbs}. D-meson candidates are defined combining triplets of tracks reconstructed in the TPC and ITS detectors with the proper charge signs, $|\eta| < 0.8$, $\pt > 0.3~\gev/c$, and a minimum of two (out of six) hits in the ITS, with at least one in either of the two innermost layers to ensure a good pointing resolution. To reduce the large combinatorial background and the contribution of \Dplusminus mesons originating from beauty-hadron decays (non-prompt), a machine-learning multi-class classification algorithm based on Boosted Decision Trees (BDT) provided by the \textsc{XGBoost} library~\cite{Chen:2016XST,barioglio_luca_2021_5070132} is employed. The variables utilized for the candidate selection in the BDT are based on the displaced decay-vertex topology, exploiting the mean proper decay length of \Dplusminus mesons of $c\tau\approx 312~\mum$~\cite{Zyla:2020zbs}, and on the PID of charged pions and kaons. Before that, a preselection of the \Dplusminus candidates based on the PID information of the decay products is applied by requiring a $3\sigma$ compatibility either with the \TPC or the \TOF expected signals of the daughter tracks. Signal samples of prompt (originating from charm-quark hadronization or decays of excited charm states) and non-prompt \Dplusminus mesons for the BDT training are obtained from MC simulations.
The background samples are obtained from the sidebands of the candidate invariant mass distributions in data. The BDT outputs are related to the candidate probability to be a prompt or non-prompt D meson, or combinatorial background. D-meson candidates are selected in the \pT interval between 1 and 10 $\gev/c$ by requiring a high probability to be a prompt \Dplusminus meson and a low probability to be a combinatorial-background candidate.

\begin{figure}[!tb]
\centering
\includegraphics[width=0.48\textwidth]{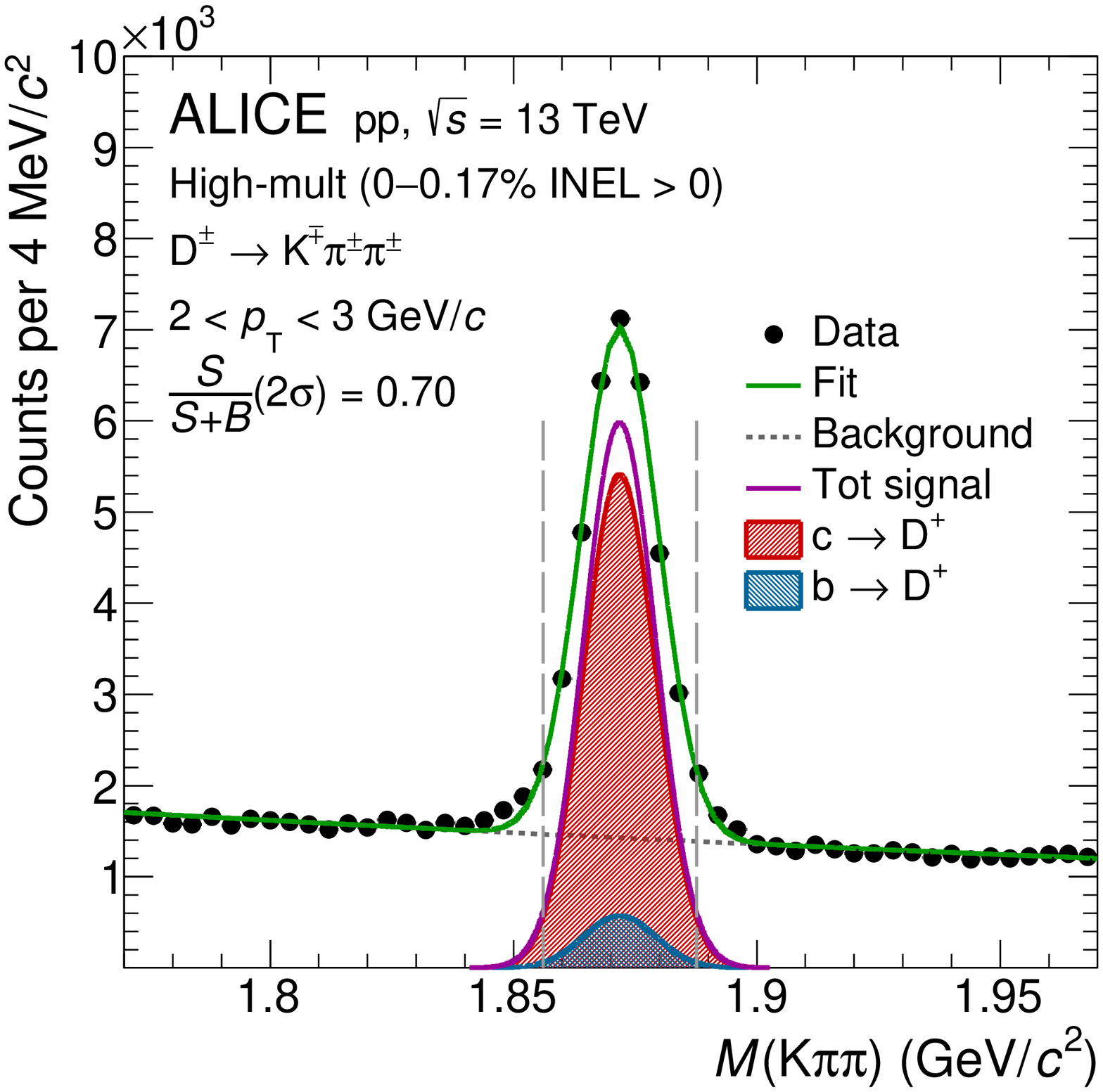}
\includegraphics[width=0.48\textwidth]{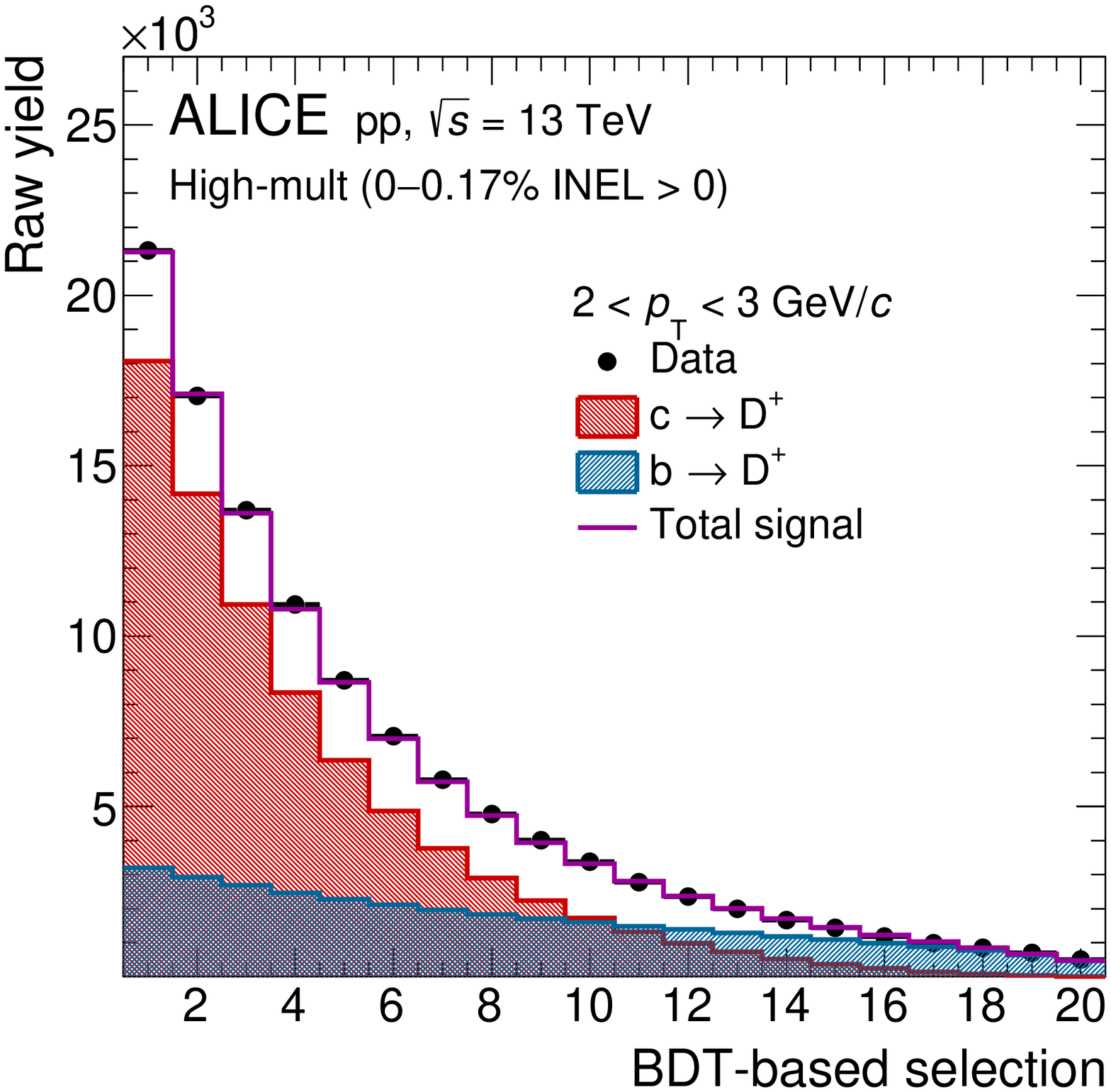}
\caption{Left: invariant mass distributions of \Dplusminus candidates in the $2<\pT<3~\gev/c$ interval. The green solid line shows the total fit function and the gray dotted line the combinatorial background. The contributions of \Dplusminus mesons originating from charm hadronisation and beauty-hadron decays are obtained with the method relying on the definition of different selection criteria, as explained in the text. Right: example of raw-yield distribution as a function of the BDT-based selection employed in the procedure adopted for the determination of the fraction of \Dplusminus originating from beauty-hadron decays for the $2<\pT<3~\gev/c$ interval.}
\label{fig:DmassAndFrac}
\end{figure}

A selection on the candidate invariant mass $(M(\mathrm{K\uppi\uppi}))$ is applied to obtain a high-purity sample of \Dplusminus mesons. To this end, the $M(\mathrm{K\uppi\uppi})$ distribution of \Dplusminus candidates is fitted in intervals of \pT of $1~\gev$ width in the range $1<\pT<10~\gev/c$ with a Gaussian function for the signal and an exponential term for the background. The left panel of Fig.~\ref{fig:DmassAndFrac} shows the $M(\mathrm{K}\uppi\uppi)$ distribution for \Dplusminus with $2<\pT<3~\gev/c$. The width of the Gaussian function used to describe the signal peak, $\sigma_{\Dplusminus}$, increases from 6 to 10 $\mev/c^2$ with increasing \pT as a consequence of the \pT dependence of the momentum resolution. The \Dplusminus-meson candidates in the invariant mass window $|M(\mathrm{K}\uppi\uppi)| < 2\sigma_{\Dplusminus}$ are selected to be paired with proton candidates. This selection, displayed by the two vertical lines in Fig.~\ref{fig:DmassAndFrac}, leads to a purity that is $P_{\Dminus} = (61.7 \pm 0.9(\mathrm{stat}) \pm 0.7(\mathrm{syst}))\%$ on average. The systematic uncertainty of $P_{\Dminus}$ is evaluated by repeating the invariant mass fits, varying the background fit function and the invariant mass upper and lower limits.

The contributions of prompt and non-prompt \Dplusminus mesons are depicted in the left panel of Fig.~\ref{fig:DmassAndFrac} with the red and blue distributions, respectively, They are obtained with a data-driven method based on the sampling of the raw yield at different values of the BDT output score related to the probability of being a non-prompt \Dplusminus meson~\cite{ALICE:2021mgk}. The yields of prompt and non-prompt \Dplusminus mesons can be extracted by solving a system of equations that relate the raw yield value $\rawY{i}$ (obtained with the $i$-th threshold on the BDT output score) to the corrected yields of prompt ($\Np$) and non-prompt ($\Nnp$) \Dplusminus mesons via the corresponding acceptance-times-efficiency factors for prompt ($\effP{i}$) and non-prompt ($\effNP{i}$) \Dplusminus mesons as follows
\begin{equation}
\renewcommand\arraystretch{1.3} {
\left(
\begin{array}{cc}
\effP{1} & \effNP{1}\\
\vdots & \vdots\\
\effP{\mathrm{n}} & \effNP{\mathrm{n}}
\end{array}
 \right)} 
 \times
 \left(
\begin{array}{c}
 \Np\\
 \Nnp
\end{array}
 \right) 
 -
  \left(
\begin{array}{c}
\rawY{1}\\
\vdots\\
Y_\mathrm{n}
\end{array}
 \right)
 = 
\left(
\begin{array}{c}
\delta_1\\
\vdots\\
\delta_\mathrm{n}
\end{array}
 \right)
.
\label{eq:systemCutVar}
\end{equation}
The $\delta_i$ factors represent the residuals that account for the equations not holding exactly due to the uncertainty of $\rawY{i}$, $\effNP{i}$, and $\effP{i}$. The system of equations can be solved via a $\chi^2$ minimization, which leads to the determination of $\Np$ and $\Nnp$. The right panel of Fig.~\ref{fig:DmassAndFrac} shows an example of a raw-yield distribution as a function of the BDT-based selection used in the minimization procedure for \Dplusminus mesons with $2<\pt<3~\gev/c$. The leftmost data point of the distribution represents the raw yield corresponding to the loosest selection on the BDT output related to the candidate probability of being a non-prompt \Dplusminus meson, while the rightmost one corresponds to the strictest selection, which is expected to preferentially select non-prompt \Dplusminus mesons. The prompt and non-prompt components, obtained for each BDT-based selection using the procedure described above, are represented by the red and blue filled histograms, respectively, while their sum is reported by the magenta histogram.

The obtained corrected yields $\Np$ and $\Nnp$ can then be used to compute the fraction of non-prompt \Dplusminus mesons $\fnonprompt^{j}$ for a given selection $j$,

\begin{equation}
\label{eq:fnpromptSystem}
  \fnonprompt^{j} = \frac{\effNP{j} \times \Nnp}{\effNP{j} \times \Nnp+\effP{j} \times \Np}.
\end{equation}

The \fnonprompt factor for the selections used to build the \pD and \apaD pairs is estimated to be $(7.7 \pm 0.5(\mathrm{stat}) \pm 0.2(\mathrm{syst}))\%$. The systematic uncertainty of \fnonprompt is evaluated by repeating the procedure with different sets of selection criteria and varying the fitting parameters in the raw-yield extraction. In addition, since the efficiency depends on the charged-particle multiplicity, the multiplicity distribution in the MC sample used for the efficiency computation was weighted in order to reproduce the one in data.

Differently from the component originating from beauty-hadron weak decays, \Dplusminus mesons originating from excited charm-meson strong decays cannot be experimentally resolved from promptly produced \Dplusminus mesons due to their short lifetime. The two largest sources are the \DstartoDpluspi and \DstartoDplusGamma decays, having $\mathrm{BR}=(30.7\pm0.5)\%$ and $\mathrm{BR}=(1.6\pm0.4)\%$~\cite{Zyla:2020zbs}, respectively. Their contribution is estimated from the production cross sections of \Dplus and \Dstarplus mesons measured in pp collisions at $\roots=5.02~\tev$~\cite{ALICE:2021mgk,ALICE:2019nxm} and employing the \textsc{Pythia~8} decayer for the description of the \DstartoDplus decay kinematics. The fraction of \Dplusminus mesons in $1<\pT<10~\gev/c$ originating from \Dstarplusminus decays is estimated to be $f_{\Dstarminus} = (27.6 \pm 1.3(\mathrm{stat}) \pm 2.4(\mathrm{syst}))\%$, where statistical and systematic uncertainties are propagated from the measurements of the \Dplus and \Dstarplus production cross sections.

\subsection{The correlation function}
\begin{figure}[tb]
\centering
\includegraphics[width=0.48\textwidth]{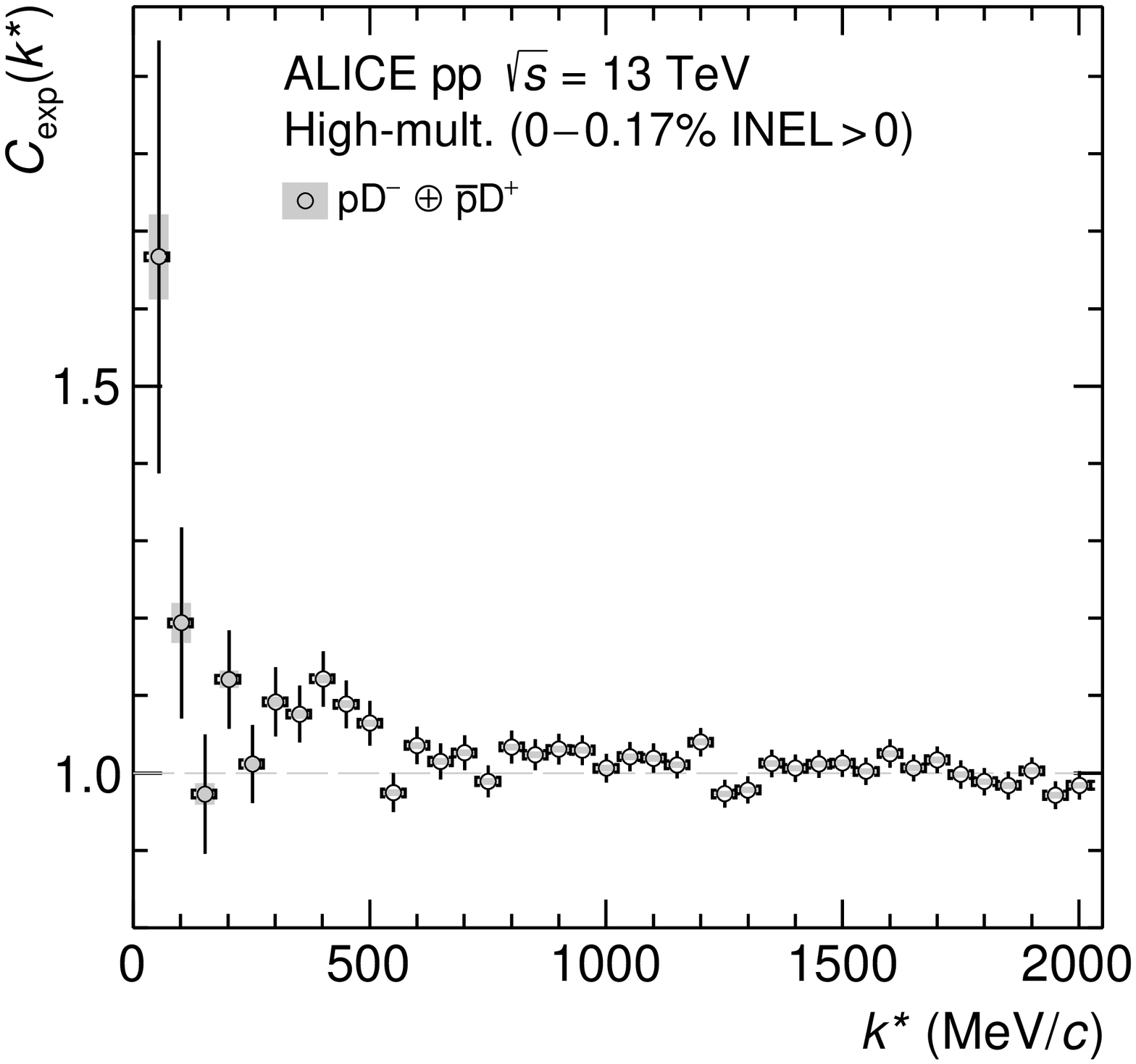}
\includegraphics[width=0.48\textwidth]{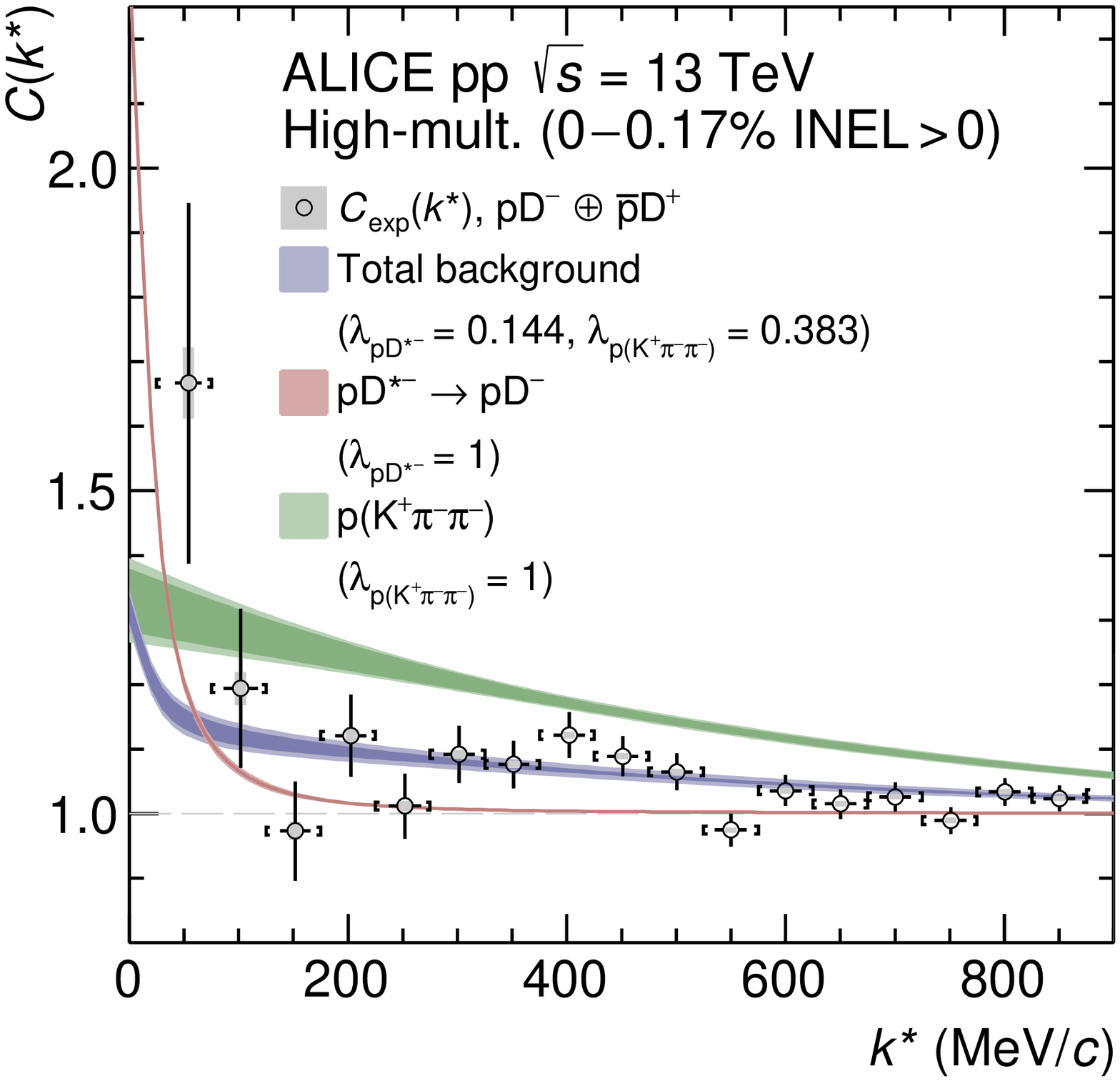}
\caption{Left: experimental \pD correlation function in the range $0<\kstar<2~\gev/c$. Statistical (bars) and systematic uncertainties (shaded boxes) are shown separately. The open boxes represent the bin width. Right: experimental \pD correlation function in a reduced \kstar range together with the contributions from \pKpipi (green band) and \pDstar (red band), and the total background model (purple band). The \pKpipi and \pDstar contributions are not scaled by the respective $\lambda$ parameter. The width of the dark (light) shaded bands depicts the statistical (total) uncertainty of the parametrized background contributions.}
\label{fig:bkg}
\end{figure}

The proton and \Dminus candidates are then combined and their relative momentum \kstar is evaluated as $\kstar=\frac{1}{2}\times|\pmb{p}^*_\mathrm{p}-\pmb{p}^*_\mathrm{D}|$, where $\pmb{p}^*_{\mathrm{p},\mathrm{D}}$ are the momenta of the two particles in the pair rest frame. The \kstar distribution of \pD pairs, $N_{\rm{same}}(\kstar)$, is then divided by the one obtained combining proton and \Dminus candidates from different events, $N_{\rm{mixed}}(\kstar)$, to compute the two-particle momentum correlation function, which is defined as $C_\mathrm{exp}(\kstar)=\mathcal{N}\times N_{\rm{same}}(\kstar)/N_{\rm{mixed}}(\kstar)$~\cite{Lisa:2005dd}. The latter provides a correction for the acceptance of the detector and the normalization for the phase space of the particle pairs. To ensure the same geometrical acceptance as for $N_{\rm{same}}$, the mixing procedure is conducted only between particle pairs produced in events with similar $z$ position of the primary vertex and similar charged-particle multiplicity. Since the correlation functions for \pD and \apaD are consistent with each other within statistical uncertainties, they are combined and in the following \pD will represent \pDcomb. The normalization constant $\mathcal{N}$ is obtained from $\kstar \in [1500, 2000]$~\si{\MeVc} where the correlation function is independent of \kstar, as expected since in this region of \kstar the pairs of particles are not affected by any interaction. The resulting correlation function $C_\mathrm{exp}(\kstar)$ is displayed in the left panel of Fig.~\ref{fig:bkg}. The data are compatible with unity for $\kstar>\SI{500}{\MeVc}$, while they show a possible hint of an increase for lower \kstar values. In total 200 \pD and 221 \apaD pairs contribute to $N_{\rm{same}}(\kstar)$ in the region of $\kstar < \SI{200}{\MeVc}$, where model calculations~\cite{Haidenbauer,Yamaguchi,Hofmann:2005sw,Fontoura} predict a deviation from unity. The systematic uncertainties of $C_\mathrm{exp}(\kstar)$ are assessed by varying the proton and \Dminus selection criteria. 

The measured two-particle momentum correlation function can be related to the source function and the two-particle wave function via the Koonin--Pratt equation $C(\kstar)=\int\mathrm{d}^3\rstar S(\rstar)|\Psi(\kstar,\rstar)|^2$~\cite{Lisa:2005dd}, where $S(\rstar)$ is the source function, $\Psi(\kstar,\rstar)$ is the two-particle wave function, and $\rstar$ refers to the relative distance between the two particles.
The source function for the \pD pairs is estimated by employing the hypothesis of a common source for all hadrons in high-multiplicity pp collisions at the LHC corrected for strong decays of extremely short-lived resonances ($c\tau\lesssim 5$~ fm) feeding into the particle pairs~\cite{FemtoSource}. This is the case for resonances strongly decaying into protons. In contrast, both beauty-hadron and \Dstarplusminus decays occur at larger distances than the typical range for the strong interaction~\cite{Zyla:2020zbs}. This implies that the correlation function for \Dminus mesons originating from these decays will only carry the imprint of the interaction of the parent particle with the proton without impacting the size of the emitting source. The core source determined in~\cite{FemtoSource} features a dependence on the transverse mass \mt of the particle pair, which can be attributed to a collective expansion of the system~\cite{Lisa:2005dd, Bearden:2000ex,Kisiel:2014upa, Shapoval:2014wya}. The collective behavior has been studied in high-multiplicity pp collisions by the CMS Collaboration and found to be comparable for light-flavor and charm hadrons~\cite{CMSCharmV2}. 
Hence, the core source of \pD pairs with $\kstar < \SI{200}{\MeVc}$ is estimated by parameterizing the measured \mt dependence of the source radius extracted from \pP correlations in~\cite{FemtoSource} and evaluating it at the $\langle\mt\rangle = \SI{2.7}{\GeVcc}$ of the \pD pairs. 
Since the production mechanism of charm mesons might not be identical to that of light-flavor baryons, the emission of the \pP and \pD pairs is studied by simulating pp collisions with \textsc{Pythia}~8.301~\cite{Sjostrand:2014zea} and computing their relative distance in the pair rest frame, $r^*$, considering only \Dminus mesons originating directly from charm-quark hadronization. These studies indicate that the core source of \pD at the pertinent $\langle\mt\rangle$ is smaller by about 25\% compared to that of \pP pairs. This is included in the systematic uncertainty of the source radius.
The resulting overall source is parametrized by a Gaussian profile characterized by an effective radius $\reff = 0.89 ^{+0.08} _{-0.22}~\si{fm}$, where the uncertainty includes both the one arising from the \mt-dependent parametrization and the \textsc{Pythia}~8 study.

The correlation function due to the genuine \pD interaction can be extracted from the measured $C_\mathrm{exp}(\kstar)$ by estimating and subtracting the contributions of \Dminus mesons originating from beauty-hadron and \Dstarminus decays, protons originating from strange-hadron decays, as well as misidentified protons and combinatorial-background D-meson candidates. The experimental correlation function is decomposed as
\begin{equation}
    C_{\rm{exp}}(\kstar) = \lambda_\mathrm{p\Dminus} \times C_\mathrm{p\Dminus}(\kstar) + \lambda_\mathrm{p(K^{+}\uppi^{-}\uppi^{-})} \times C_{\mathrm{p(K}^{+}\uppi^{-}\uppi^{-})}(\kstar) + \lambda_\mathrm{p\Dstarminus} \times C_\mathrm{p\Dstarminus}(\kstar) + \lambda_{\mathrm{flat}} \times C_{\mathrm{flat}}.
    \label{eq:CFexp}
\end{equation}
The combinatorial \Kpipi background below the \Dminus peak and the final-state interaction among protons and \Dminus from \Dstarminus decays play a significant role. All other contributions are assumed to be characterized by a $C(\kstar)$ compatible with unity and are therefore included in the $C_\mathrm{flat}$ contribution. The relative weights, $\lambda_i$, are evaluated considering the contributions to \Dminus candidates described above and following the procedure explained in~\cite{ALICE:2018ysd} for the protons. They are about 33.9\% for $C_\mathrm{p\Dminus}(\kstar)$ and 38.8\%, 14.4\%, and 13.4\% for the \pKpipi, \pDstar, and flat contributions, respectively.

The correlation function $C_{\mathrm{p(K}^{+}\uppi^{-}\uppi^{-})}$ is extracted from the sidebands of the \Dminus candidates, chosen as $[M_{\Dminus}(\pT) - \SI{200}{\MeVc},~M_{\Dminus}(\pT) - 5 \times \sigma_{\Dminus}(\pT)]$ and $[M_{\Dminus}(\pT) + 5 \times \sigma_{\Dminus}(\pT),~M_{\Dminus}(\pT) + \SI{200}{\MeVc}]$ for the left and right sidebands, respectively.
The contamination from $\DstarminustoDzeropi$ decays in the right sideband is suppressed by a $2.5\sigma_{\Dstarminus}$ rejection around the mean value of the \Dstarminus invariant mass peak. The resulting correlation function is parametrized by a third-order polynomial in $\kstar\in[0,~1.5]~\si{\GeVc}$ and is displayed by the green curve reported in the right panel of Fig.~\ref{fig:bkg}. The observed behavior is determined by meson--meson and baryon--meson mini-jets and residual two-body interactions among the quadruplet, as obtained from previous studies~\cite{FemtopSigma, FemtopPhi}.

The residual \pDstar correlation function is computed employing the Koonin--Pratt formalism using the CATS framework~\cite{Mihaylov:2018rva} 
to obtain a two-particle wave function $\Psi(\kstar,\rstar)$ considering only the Coulomb interaction and assuming that the source radius is the same as for \pD pairs. The obtained \pDstar correlation function is transformed to the momentum basis of the \pD relative momentum by considering the kinematics of the \DstartoDminus decay~\cite{PhysRevC.89.054916}. The resulting correlation function is shown in the right panel of Fig.~\ref{fig:bkg} as a red band. The purple band in the same figure represents the total background that includes all contributions with their corresponding weights.
Finally, the genuine \pD correlation function is obtained by solving Eq.~\ref{eq:CFexp} for $C_\mathrm{p\Dminus}(\kstar)$ and is shown in Fig.~\ref{fig:result}.

The systematic uncertainties of the genuine \pD correlation function, $C_\mathrm{p\Dminus}(\kstar)$, include (i) the uncertainties of $C_\mathrm{exp}(\kstar)$, (ii) the uncertainties of the $\lambda_i$ weights, and (iii) the uncertainties related to the parametrization of the background sources, $C_\mathrm{p(K^{+}\uppi^{-}\uppi^{-})}(\kstar)$ and $C_\mathrm{p\Dstarminus}(\kstar)$. In particular, as previously mentioned, the systematic uncertainty on $C_\mathrm{exp}(\kstar)$ is estimated by varying the proton and \Dminus-candidate selection criteria and ranges between 0.5\% and 3\% as a function of \kstar. The uncertainties of the $\lambda_i$ weights are derived from the systematic uncertainties on the proton and \Dminus purities ($P_\mathrm{p}$ and $P_\mathrm{\Dminus}$), $f_{\Dstarminus}$, and \fnonprompt reported in Section~\ref{sec:selection}. The systematic uncertainties of $C_\mathrm{p(K^{+}\uppi^{-}\uppi^{-})}(\kstar)$ are estimated following the same procedure adopted for $C_\mathrm{exp}(\kstar)$ and, in addition, by varying the range of the fit of the correlation function parametrized from the sidebands regions of the invariant mass distribution. Additional checks are performed by varying the invariant mass interval used to define the sidebands region of up to 100 $\mev/c^2$ . The resulting systematic uncertainty ranges from 1\% to 5\%. The systematic uncertainty of $C_\mathrm{p\Dstarminus}(\kstar)$ is due to the uncertainty on the emitting source. Considering the small $\lambda_\mathrm{p\Dstarminus}(\kstar)$ this uncertainty results to be negligible compared to the other sources of uncertainty. The overall relative systematic uncertainty on $C_\mathrm{p\Dminus}(\kstar)$ resulting from the different sources ranges between 3\% and 10\% and is maximum in the lowest \kstar interval.

\section{Results}
The resulting genuine $C_\mathrm{p\Dminus}(\kstar)$ correlation function can be employed to study the \pD strong interaction that is characterized by two isospin configurations and is coupled to the \nDbar channel. First of all, in order to assess the effect of the strong interaction on the correlation function, a reference calculation including only the Coulomb interaction is considered. The corresponding correlation function is obtained using CATS~\cite{Mihaylov:2018rva}.
Secondly, various theoretical approaches to describe the strong interaction are benchmarked, including meson exchange (J. Haidenbauer \etal.~\cite{Haidenbauer}), meson exchange based on heavy quark symmetry (Y. Yamaguchi \etal.~\cite{Yamaguchi}), an SU(4) contact interaction (J. Hoffmann and M. Lutz~\cite{Hofmann:2005sw}), and a chiral quark model (C. Fontoura \etal.~\cite{Fontoura}).
The relative wave functions for the model of J. Haidenbauer \etal.~\cite{Haidenbauer} are provided directly, while for the other models~\cite{Hofmann:2005sw,Yamaguchi,Fontoura} they are evaluated by employing a Gaussian potential whose strength is adjusted to describe the corresponding published $\mathrm{I}=0$ and $\mathrm{I}=1$ scattering lengths listed in Table~\ref{tab:scatParam}. The \pD correlation function is computed within the Koonin--Pratt formalism, taking into account explicitly the coupling between the \pD and \nDbar channels~\cite{Lednicky:1998} and including the Coulomb interaction~\cite{Kamiya:2019uiw}. The finite experimental momentum resolution is considered in the modeling of the correlation functions~\cite{ALICE:2018ysd}.

\begin{figure}[!tb]
\centering
\includegraphics[width=0.5\textwidth]{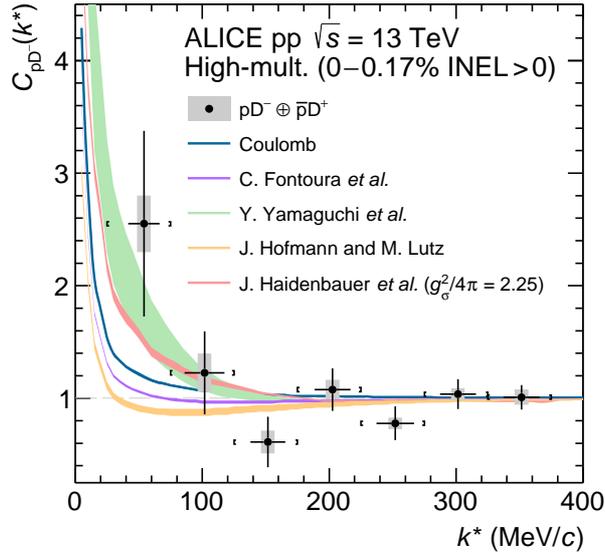}
\caption{Genuine \pD correlation function  compared with different theoretical models (see text for details). The null hypothesis is represented by the curve corresponding to the Coulomb interaction only.}
\label{fig:result}
\end{figure}

The outcome of these models is compared in Fig.~\ref{fig:result} with the measured genuine \pD correlation function.
The degree of consistency between data and models is quantified by the p-value computed in the range $\kstar < \SI{200}{\MeVc}$. It is expressed by the number of standard deviations $n_\sigma$ reported in Table~\ref{tab:scatParam}, where the $n_\sigma$ range accounts, at one standard deviation level, for the total uncertainties of the data points and the models. The values of the scattering lengths $f_0$ for the different models are also reported in Table~\ref{tab:scatParam}. Here, the high-energy physics convention on the scattering-length sign is adopted: a negative value corresponds to either a repulsive interaction or to an attractive one with presence of a bound state, while a positive value corresponds to an attractive interaction. The data are compatible with the Coulomb-only hypothesis within $\NsigmaCoulomb\sigma$. Nevertheless, the level of agreement slightly improves in case of the models by
J. Haidenbauer \etal. (employing $g_{\sigma}^2/4\pi = 2.25$) which predicts an attractive interaction, and by Y. Yamaguchi \etal. which foresees the formation of a \DbarN bound state with a mass of \SI{2804}{\MeVcc} in the $\mathrm{I}=0$ channel.

\begin{table}[!b]
\caption{Scattering parameters of the different theoretical models for the \DbarN interaction~\cite{Haidenbauer, Hofmann:2005sw, Yamaguchi, Fontoura} and degree of consistency with the experimental data computed in the range $\kstar < \SI{200}{\MeVc}$.
}
\centering\begin{tabular}{l | c c | c}
Model & $f_0~(\mathrm{I}=0)$ & $f_0~(\mathrm{I}=1)$ & $n_\sigma$ \\
\hline
Coulomb & & & \NsigmaCoulomb \\
Haidenbauer \etal.~\cite{Haidenbauer} & $0.67$ & $0.04$ & \NsigmaHaidenbauerMod \\
~ ($g_{\sigma}^2/4\pi = 2.25)$ & & \\ 
Hofmann and Lutz~\cite{Hofmann:2005sw} & $-0.16$ & $-0.26$ & \NsigmaHofmannLutz \\
Yamaguchi \etal.~\cite{Yamaguchi} & $-4.38$ & $-0.07$ & \NsigmaYamaguchi \\
Fontoura \etal.~\cite{Fontoura} & $0.16$ & $-0.25$ & \NsigmaFontouraTwo\\
\end{tabular}
\label{tab:scatParam}
\end{table}

\begin{figure}[!tb]
\centering
\includegraphics[width=0.48\textwidth]{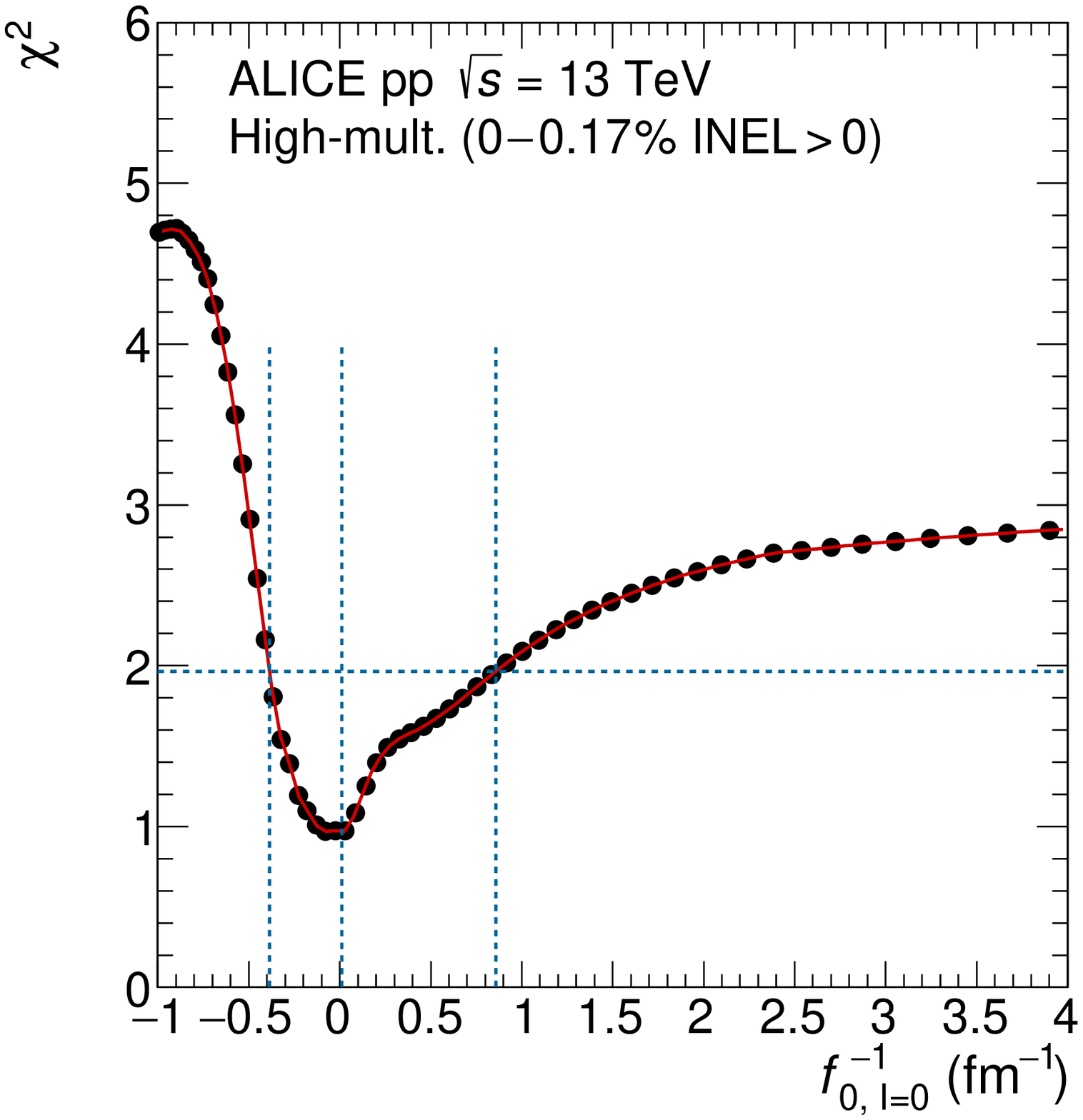}
\includegraphics[width=0.48\textwidth]{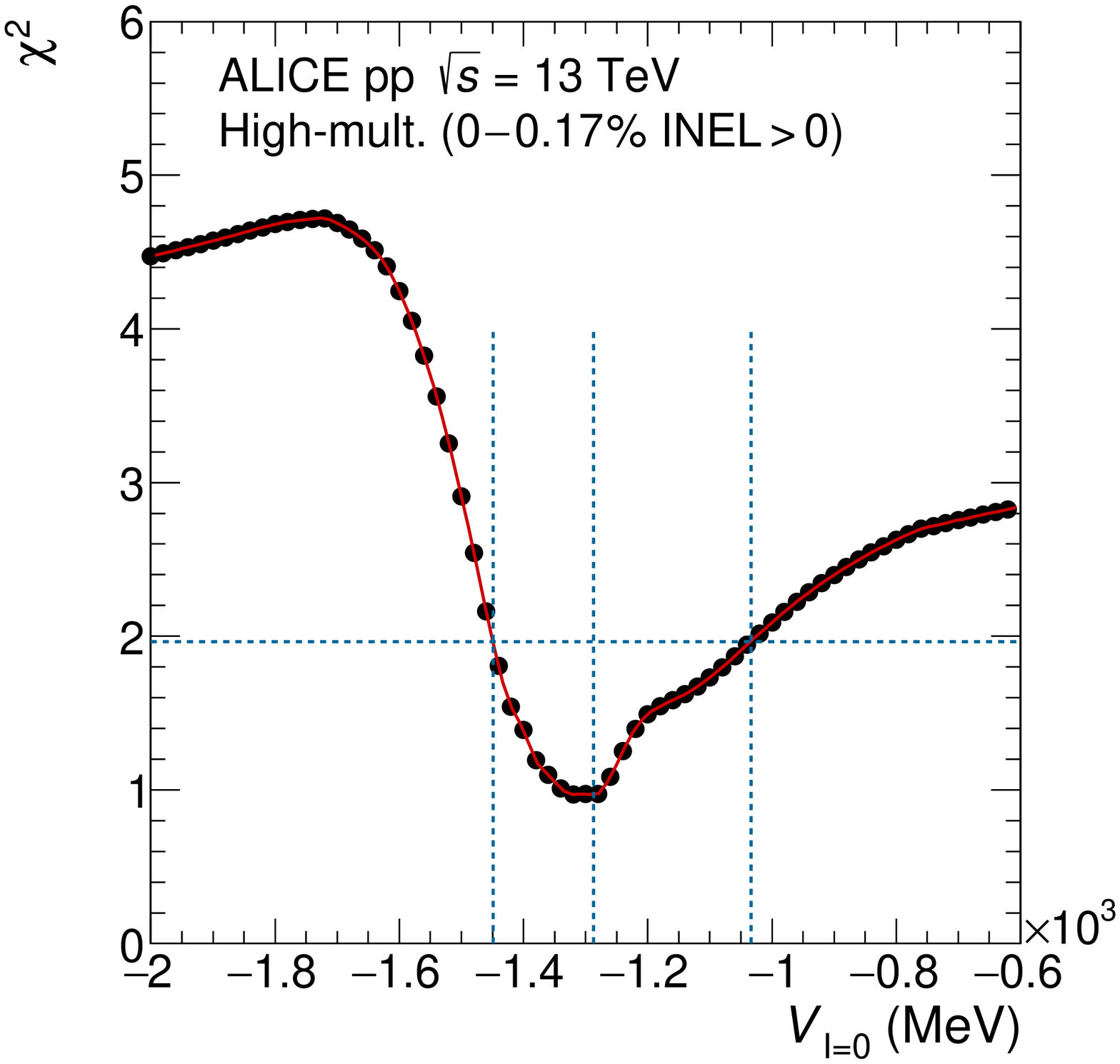}
\caption{$\chi^2$ distributions obtained by comparing the measured $C_{\pD}(\kstar)$ for $\kstar < \SI{200}{\MeVc}$ with the correlation function calculated with an interaction modeled by a Gaussian potential with an interaction range given by $\uprho$-meson exchanges as a function of the inverse scattering length (left panel) and the interaction potential (right panel) for $\mathrm{I}=0$. The blue dotted lines represent the value of $f_{0,~\mathrm{I}=0}^{-1}$ and $V_\mathrm{I}=0$ for which the $\chi^2$ is minimum and for the $1\,\sigma$ confidence interval.}
\label{fig:chi2}
\end{figure}

\begin{figure}[!tb]
\centering
\includegraphics[width=0.5\textwidth]{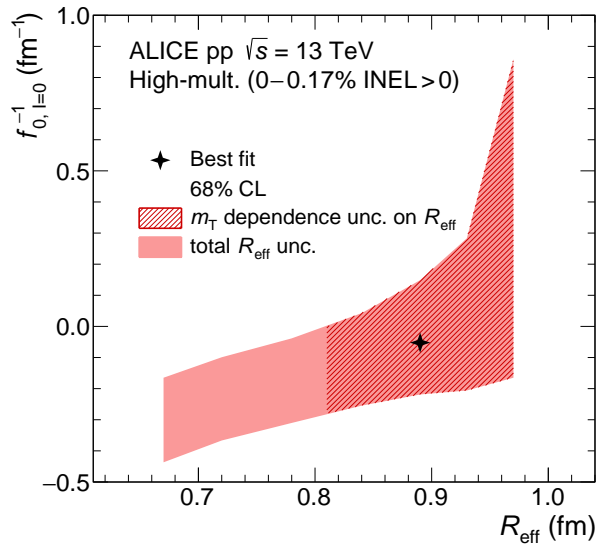}
\caption{Regions of 68\% confidence intervals for the inverse scattering length \scattLen as a function of the source radius varied within one standard deviation considering only the \mt dependence on \reff and the total uncertainty (see text for details) under the assumption of negligible interaction for $\mathrm{I}=1$. The most probable value is reported by the star symbol.}
\label{fig:CI}
\end{figure}

Finally, the scattering parameters can be constrained by comparing the data with the outcome of calculations carried out varying the strength of the potential and the source radius. In this case the interaction potential is parametrized by a Gaussian-type functional form with the range of $\uprho$-meson exchange. In this estimation, it is assumed that the interaction in the $\mathrm{I}=1$ channel is negligible for simplicity. The correlation function $C_{\pD}(\kstar)$ is computed including also the Coulomb interaction and the coupled channel. This procedure is repeated for different values of the interaction potential for the $\mathrm{I}=0$ channel ($V_\mathrm{I=0}$). For all the correlation functions corresponding to the different interaction potentials, the agreement with the data is evaluated by computing the $\chi^2$ using a bootstrap procedure. Both the statistical and systematic uncertainties of the data are considered in the bootstrap procedure, as well as the uncertainty on the emitting source radius (\reff) in the computed $C_{\pD}(\kstar)$, which is varied within $1\sigma$ of its uncertainty.
The resulting overall $\chi^2$ distributions are shown in Fig.~\ref{fig:chi2} as a function of $f_{0,~I=0}^{-1}$ and $V_\mathrm{I=0}$ in the left and right panels, respectively. The data are found to be consistent with a potential strength of \resultPot within $1\sigma$. This corresponds to an inverse scattering-length interval of \resultInvScatt. Since the determined potential strength is always attractive, the positive values of the scattering length imply an attractive interaction without bound states, while the negative values are consistent with the presence of a \DbarN bound state. The same procedure was repeated for fixed values of \reff in order to obtain the $1\sigma$ confidence interval as a function of the emitting source radius. Figure~\ref{fig:CI} shows the confidence interval as a function of the source radius varied within $1\sigma$ of its uncertainty. The dashed interval corresponds to the radius uncertainty due to only the \mt dependence while the full-shaded interval shows the total radius uncertainty. The most probable value reported in Fig~\ref{fig:CI} with the star symbol corresponds to an attractive interaction with the formation of a bound state. Given that most models predict a repulsive $\mathrm{I}=1$ interaction, in reality the $\mathrm{I}=0$ interaction might have to be even more attractive. The herewith presented limits provide valuable guidance for further theoretical studies advancing the understanding of the strong interaction in the charm sector.

\section{Summary}
In conclusion, this article presents the first measurement of correlation functions involving charm hadrons, which allows one to access to the strong interaction between a proton and a charm meson. The genuine \pD correlation function reflects the pattern of an overall attractive interaction. The data are compatible within $\NsigmaCoulomb\sigma$ with the correlation function obtained from the hypothesis of a Coulomb-only interaction. The degree of consistency improves when  considering, in addition, state-of-the-art models that predict an attractive strong \DbarN interaction with or without a bound state. Finally, assuming no interaction for the $\mathrm{I}=1$ channel, the scattering length of the \DbarN system in the isospin $\mathrm{I}=0$ channel is estimated as \resultInvScatt. 
This exploratory study paves the way for precision studies of the strong interactions involving charm hadrons, facilitated by about one order of magnitude larger pp data samples expected to be collected in the next years during the LHC Runs 3 and 4~\cite{LHCppFemto}.


\newenvironment{acknowledgement}{\relax}{\relax}
\begin{acknowledgement}
\section*{Acknowledgements}

The ALICE Collaboration would like to thank all its engineers and technicians for their invaluable contributions to the construction of the experiment and the CERN accelerator teams for the outstanding performance of the LHC complex.
The ALICE Collaboration gratefully acknowledges the resources and support provided by all Grid centres and the Worldwide LHC Computing Grid (WLCG) collaboration.
The ALICE Collaboration acknowledges the following funding agencies for their support in building and running the ALICE detector:
A. I. Alikhanyan National Science Laboratory (Yerevan Physics Institute) Foundation (ANSL), State Committee of Science and World Federation of Scientists (WFS), Armenia;
Austrian Academy of Sciences, Austrian Science Fund (FWF): [M 2467-N36] and Nationalstiftung f\"{u}r Forschung, Technologie und Entwicklung, Austria;
Ministry of Communications and High Technologies, National Nuclear Research Center, Azerbaijan;
Conselho Nacional de Desenvolvimento Cient\'{\i}fico e Tecnol\'{o}gico (CNPq), Financiadora de Estudos e Projetos (Finep), Funda\c{c}\~{a}o de Amparo \`{a} Pesquisa do Estado de S\~{a}o Paulo (FAPESP) and Universidade Federal do Rio Grande do Sul (UFRGS), Brazil;
Ministry of Education of China (MOEC) , Ministry of Science \& Technology of China (MSTC) and National Natural Science Foundation of China (NSFC), China;
Ministry of Science and Education and Croatian Science Foundation, Croatia;
Centro de Aplicaciones Tecnol\'{o}gicas y Desarrollo Nuclear (CEADEN), Cubaenerg\'{\i}a, Cuba;
Ministry of Education, Youth and Sports of the Czech Republic, Czech Republic;
The Danish Council for Independent Research | Natural Sciences, the VILLUM FONDEN and Danish National Research Foundation (DNRF), Denmark;
Helsinki Institute of Physics (HIP), Finland;
Commissariat \`{a} l'Energie Atomique (CEA) and Institut National de Physique Nucl\'{e}aire et de Physique des Particules (IN2P3) and Centre National de la Recherche Scientifique (CNRS), France;
Bundesministerium f\"{u}r Bildung und Forschung (BMBF) and GSI Helmholtzzentrum f\"{u}r Schwerionenforschung GmbH, Germany;
General Secretariat for Research and Technology, Ministry of Education, Research and Religions, Greece;
National Research, Development and Innovation Office, Hungary;
Department of Atomic Energy Government of India (DAE), Department of Science and Technology, Government of India (DST), University Grants Commission, Government of India (UGC) and Council of Scientific and Industrial Research (CSIR), India;
Indonesian Institute of Science, Indonesia;
Istituto Nazionale di Fisica Nucleare (INFN), Italy;
Japanese Ministry of Education, Culture, Sports, Science and Technology (MEXT) and Japan Society for the Promotion of Science (JSPS) KAKENHI, Japan;
Consejo Nacional de Ciencia (CONACYT) y Tecnolog\'{i}a, through Fondo de Cooperaci\'{o}n Internacional en Ciencia y Tecnolog\'{i}a (FONCICYT) and Direcci\'{o}n General de Asuntos del Personal Academico (DGAPA), Mexico;
Nederlandse Organisatie voor Wetenschappelijk Onderzoek (NWO), Netherlands;
The Research Council of Norway, Norway;
Commission on Science and Technology for Sustainable Development in the South (COMSATS), Pakistan;
Pontificia Universidad Cat\'{o}lica del Per\'{u}, Peru;
Ministry of Education and Science, National Science Centre and WUT ID-UB, Poland;
Korea Institute of Science and Technology Information and National Research Foundation of Korea (NRF), Republic of Korea;
Ministry of Education and Scientific Research, Institute of Atomic Physics, Ministry of Research and Innovation and Institute of Atomic Physics and University Politehnica of Bucharest, Romania;
Joint Institute for Nuclear Research (JINR), Ministry of Education and Science of the Russian Federation, National Research Centre Kurchatov Institute, Russian Science Foundation and Russian Foundation for Basic Research, Russia;
Ministry of Education, Science, Research and Sport of the Slovak Republic, Slovakia;
National Research Foundation of South Africa, South Africa;
Swedish Research Council (VR) and Knut \& Alice Wallenberg Foundation (KAW), Sweden;
European Organization for Nuclear Research, Switzerland;
Suranaree University of Technology (SUT), National Science and Technology Development Agency (NSDTA), Suranaree University of Technology (SUT), Thailand Science Research and Innovation (TSRI) and National Science, Research and Innovation Fund (NSRF), Thailand;
Turkish Energy, Nuclear and Mineral Research Agency (TENMAK), Turkey;
National Academy of  Sciences of Ukraine, Ukraine;
Science and Technology Facilities Council (STFC), United Kingdom;
National Science Foundation of the United States of America (NSF) and United States Department of Energy, Office of Nuclear Physics (DOE NP), United States of America.
\end{acknowledgement}

\bibliographystyle{utphys}   
\bibliography{bibliography}

\newpage
\appendix

%
%

\section{The ALICE Collaboration}
\label{app:collab}

\begingroup
\small
\begin{flushleft}
S.~Acharya\Irefn{org142}\And
D.~Adamov\'{a}\Irefn{org96}\And
A.~Adler\Irefn{org74}\And
J.~Adolfsson\Irefn{org81}\And
G.~Aglieri Rinella\Irefn{org34}\And
M.~Agnello\Irefn{org30}\And
N.~Agrawal\Irefn{org54}\And
Z.~Ahammed\Irefn{org142}\And
S.~Ahmad\Irefn{org16}\And
S.U.~Ahn\Irefn{org76}
% ...
\renewcommand\labelenumi{\textsuperscript{\theenumi}~}

\section*{Affiliation notes}
\renewcommand\theenumi{\roman{enumi}}
\begin{Authlist}
\item \Adef{orgI} Deceased
\item \Adef{orgII}Also at: Italian National Agency for New Technologies, Energy and Sustainable Economic Development (ENEA), Bologna, Italy
\item \Adef{orgIII}Also at: Dipartimento DET del Politecnico di Torino, Turin, Italy
\item \Adef{orgIV}Also at: M.V. Lomonosov Moscow State University, D.V. Skobeltsyn Institute of Nuclear, Physics, Moscow, Russia
\item \Adef{orgV}Also at: Department of Applied Physics, Aligarh Muslim University, Aligarh, India
\item \Adef{orgVI}Also at: Institute of Theoretical Physics, University of Wroclaw, Poland
\item \Adef{orgVII}Also at: University of Kansas, Lawrence, Kansas, United States
\end{Authlist}

\section*{Collaboration Institutes}
\renewcommand\theenumi{\arabic{enumi}~}
\begin{Authlist}
\item \Idef{org1} A.I. Alikhanyan National Science Laboratory (Yerevan Physics Institute) Foundation, Yerevan, Armenia
\item \Idef{org2} AGH University of Science and Technology, Cracow, Poland
\item \Idef{org3} Bogolyubov Institute for Theoretical Physics, National Academy of Sciences of Ukraine, Kiev, Ukraine
\item \Idef{org4} Bose Institute, Department of Physics  and Centre for Astroparticle Physics and Space Science (CAPSS), Kolkata, India
\item \Idef{org5} Budker Institute for Nuclear Physics, Novosibirsk, Russia
\item \Idef{org6} California Polytechnic State University, San Luis Obispo, California, United States
\item \Idef{org7} Central China Normal University, Wuhan, China
\item \Idef{org8} Centro de Aplicaciones Tecnol\'{o}gicas y Desarrollo Nuclear (CEADEN), Havana, Cuba
\item \Idef{org9} Centro de Investigaci\'{o}n y de Estudios Avanzados (CINVESTAV), Mexico City and M\'{e}rida, Mexico
\item \Idef{org10} Chicago State University, Chicago, Illinois, United States
\item \Idef{org11} China Institute of Atomic Energy, Beijing, China
\item \Idef{org12} Chungbuk National University, Cheongju, Republic of Korea
\item \Idef{org13} Comenius University Bratislava, Faculty of Mathematics, Physics and Informatics, Bratislava, Slovakia
\item \Idef{org14} COMSATS University Islamabad, Islamabad, Pakistan
\item \Idef{org15} Creighton University, Omaha, Nebraska, United States
\item \Idef{org16} Department of Physics, Aligarh Muslim University, Aligarh, India
\item \Idef{org17} Department of Physics, Pusan National University, Pusan, Republic of Korea
\item \Idef{org18} Department of Physics, Sejong University, Seoul, Republic of Korea
\item \Idef{org19} Department of Physics, University of California, Berkeley, California, United States
\item \Idef{org20} Department of Physics, University of Oslo, Oslo, Norway
\item \Idef{org21} Department of Physics and Technology, University of Bergen, Bergen, Norway
\item \Idef{org22} Dipartimento di Fisica dell'Universit\`{a} and Sezione INFN, Cagliari, Italy
\item \Idef{org23} Dipartimento di Fisica dell'Universit\`{a} and Sezione INFN, Trieste, Italy
\item \Idef{org24} Dipartimento di Fisica dell'Universit\`{a} and Sezione INFN, Turin, Italy
\item \Idef{org25} Dipartimento di Fisica e Astronomia dell'Universit\`{a} and Sezione INFN, Bologna, Italy
\item \Idef{org26} Dipartimento di Fisica e Astronomia dell'Universit\`{a} and Sezione INFN, Catania, Italy
\item \Idef{org27} Dipartimento di Fisica e Astronomia dell'Universit\`{a} and Sezione INFN, Padova, Italy
\item \Idef{org28} Dipartimento di Fisica e Nucleare e Teorica, Universit\`{a} di Pavia, Pavia, Italy
\item \Idef{org29} Dipartimento di Fisica `E.R.~Caianiello' dell'Universit\`{a} and Gruppo Collegato INFN, Salerno, Italy
\item \Idef{org30} Dipartimento DISAT del Politecnico and Sezione INFN, Turin, Italy
\item \Idef{org31} Dipartimento di Scienze e Innovazione Tecnologica dell'Universit\`{a} del Piemonte Orientale and INFN Sezione di Torino, Alessandria, Italy
\item \Idef{org32} Dipartimento di Scienze MIFT, Universit\`{a} di Messina, Messina, Italy
\item \Idef{org33} Dipartimento Interateneo di Fisica `M.~Merlin' and Sezione INFN, Bari, Italy
\item \Idef{org34} European Organization for Nuclear Research (CERN), Geneva, Switzerland
\item \Idef{org35} Faculty of Electrical Engineering, Mechanical Engineering and Naval Architecture, University of Split, Split, Croatia
\item \Idef{org36} Faculty of Engineering and Science, Western Norway University of Applied Sciences, Bergen, Norway
\item \Idef{org37} Faculty of Nuclear Sciences and Physical Engineering, Czech Technical University in Prague, Prague, Czech Republic
\item \Idef{org38} Faculty of Science, P.J.~\v{S}af\'{a}rik University, Ko\v{s}ice, Slovakia
\item \Idef{org39} Frankfurt Institute for Advanced Studies, Johann Wolfgang Goethe-Universit\"{a}t Frankfurt, Frankfurt, Germany
\item \Idef{org40} Fudan University, Shanghai, China
\item \Idef{org41} Gangneung-Wonju National University, Gangneung, Republic of Korea
\item \Idef{org42} Gauhati University, Department of Physics, Guwahati, India
\item \Idef{org43} Helmholtz-Institut f\"{u}r Strahlen- und Kernphysik, Rheinische Friedrich-Wilhelms-Universit\"{a}t Bonn, Bonn, Germany
\item \Idef{org44} Helsinki Institute of Physics (HIP), Helsinki, Finland
\item \Idef{org45} High Energy Physics Group,  Universidad Aut\'{o}noma de Puebla, Puebla, Mexico
\item \Idef{org46} Hiroshima University, Hiroshima, Japan
\item \Idef{org47} Hochschule Worms, Zentrum  f\"{u}r Technologietransfer und Telekommunikation (ZTT), Worms, Germany
\item \Idef{org48} Horia Hulubei National Institute of Physics and Nuclear Engineering, Bucharest, Romania
\item \Idef{org49} Indian Institute of Technology Bombay (IIT), Mumbai, India
\item \Idef{org50} Indian Institute of Technology Indore, Indore, India
\item \Idef{org51} Indonesian Institute of Sciences, Jakarta, Indonesia
\item \Idef{org52} INFN, Laboratori Nazionali di Frascati, Frascati, Italy
\item \Idef{org53} INFN, Sezione di Bari, Bari, Italy
\item \Idef{org54} INFN, Sezione di Bologna, Bologna, Italy
\item \Idef{org55} INFN, Sezione di Cagliari, Cagliari, Italy
\item \Idef{org56} INFN, Sezione di Catania, Catania, Italy
\item \Idef{org57} INFN, Sezione di Padova, Padova, Italy
\item \Idef{org58} INFN, Sezione di Pavia, Pavia, Italy
\item \Idef{org59} INFN, Sezione di Torino, Turin, Italy
\item \Idef{org60} INFN, Sezione di Trieste, Trieste, Italy
\item \Idef{org61} Inha University, Incheon, Republic of Korea
\item \Idef{org149} Institute for Advanced Simulation, Forschungszentrum J\"ulich, J\"ulich, Germany
\item \Idef{org62} Institute for Gravitational and Subatomic Physics (GRASP), Utrecht University/Nikhef, Utrecht, Netherlands
\item \Idef{org63} Institute for Nuclear Research, Academy of Sciences, Moscow, Russia
\item \Idef{org64} Institute of Experimental Physics, Slovak Academy of Sciences, Ko\v{s}ice, Slovakia
\item \Idef{org65} Institute of Physics, Homi Bhabha National Institute, Bhubaneswar, India
\item \Idef{org66} Institute of Physics of the Czech Academy of Sciences, Prague, Czech Republic
\item \Idef{org67} Institute of Space Science (ISS), Bucharest, Romania
\item \Idef{org68} Institut f\"{u}r Kernphysik, Johann Wolfgang Goethe-Universit\"{a}t Frankfurt, Frankfurt, Germany
\item \Idef{org69} Instituto de Ciencias Nucleares, Universidad Nacional Aut\'{o}noma de M\'{e}xico, Mexico City, Mexico
\item \Idef{org70} Instituto de F\'{i}sica, Universidade Federal do Rio Grande do Sul (UFRGS), Porto Alegre, Brazil
\item \Idef{org71} Instituto de F\'{\i}sica, Universidad Nacional Aut\'{o}noma de M\'{e}xico, Mexico City, Mexico
\item \Idef{org72} iThemba LABS, National Research Foundation, Somerset West, South Africa
\item \Idef{org73} Jeonbuk National University, Jeonju, Republic of Korea
\item \Idef{org74} Johann-Wolfgang-Goethe Universit\"{a}t Frankfurt Institut f\"{u}r Informatik, Fachbereich Informatik und Mathematik, Frankfurt, Germany
\item \Idef{org75} Joint Institute for Nuclear Research (JINR), Dubna, Russia
\item \Idef{org76} Korea Institute of Science and Technology Information, Daejeon, Republic of Korea
\item \Idef{org77} KTO Karatay University, Konya, Turkey
\item \Idef{org78} Laboratoire de Physique des 2 Infinis, Ir\`{e}ne Joliot-Curie, Orsay, France
\item \Idef{org79} Laboratoire de Physique Subatomique et de Cosmologie, Universit\'{e} Grenoble-Alpes, CNRS-IN2P3, Grenoble, France
\item \Idef{org80} Lawrence Berkeley National Laboratory, Berkeley, California, United States
\item \Idef{org81} Lund University Department of Physics, Division of Particle Physics, Lund, Sweden
\item \Idef{org82} Moscow Institute for Physics and Technology, Moscow, Russia
\item \Idef{org83} Nagasaki Institute of Applied Science, Nagasaki, Japan
\item \Idef{org84} Nara Women{'}s University (NWU), Nara, Japan
\item \Idef{org85} National and Kapodistrian University of Athens, School of Science, Department of Physics , Athens, Greece
\item \Idef{org86} National Centre for Nuclear Research, Warsaw, Poland
\item \Idef{org87} National Institute of Science Education and Research, Homi Bhabha National Institute, Jatni, India
\item \Idef{org88} National Nuclear Research Center, Baku, Azerbaijan
\item \Idef{org89} National Research Centre Kurchatov Institute, Moscow, Russia
\item \Idef{org90} Niels Bohr Institute, University of Copenhagen, Copenhagen, Denmark
\item \Idef{org91} Nikhef, National institute for subatomic physics, Amsterdam, Netherlands
\item \Idef{org92} NRC Kurchatov Institute IHEP, Protvino, Russia
\item \Idef{org93} NRC \guillemotleft Kurchatov\guillemotright  Institute - ITEP, Moscow, Russia
\item \Idef{org94} NRNU Moscow Engineering Physics Institute, Moscow, Russia
\item \Idef{org95} Nuclear Physics Group, STFC Daresbury Laboratory, Daresbury, United Kingdom
\item \Idef{org96} Nuclear Physics Institute of the Czech Academy of Sciences, \v{R}e\v{z} u Prahy, Czech Republic
\item \Idef{org97} Oak Ridge National Laboratory, Oak Ridge, Tennessee, United States
\item \Idef{org98} Ohio State University, Columbus, Ohio, United States
\item \Idef{org99} Petersburg Nuclear Physics Institute, Gatchina, Russia
\item \Idef{org100} Physics department, Faculty of science, University of Zagreb, Zagreb, Croatia
\item \Idef{org101} Physics Department, Panjab University, Chandigarh, India
\item \Idef{org102} Physics Department, University of Jammu, Jammu, India
\item \Idef{org103} Physics Department, University of Rajasthan, Jaipur, India
\item \Idef{org104} Physikalisches Institut, Eberhard-Karls-Universit\"{a}t T\"{u}bingen, T\"{u}bingen, Germany
\item \Idef{org105} Physikalisches Institut, Ruprecht-Karls-Universit\"{a}t Heidelberg, Heidelberg, Germany
\item \Idef{org106} Physik Department, Technische Universit\"{a}t M\"{u}nchen, Munich, Germany
\item \Idef{org107} Politecnico di Bari and Sezione INFN, Bari, Italy
\item \Idef{org108} Research Division and ExtreMe Matter Institute EMMI, GSI Helmholtzzentrum f\"ur Schwerionenforschung GmbH, Darmstadt, Germany
\item \Idef{org151} RIKEN iTHEMS, Wako, Japan
\item \Idef{org109} Russian Federal Nuclear Center (VNIIEF), Sarov, Russia
\item \Idef{org110} Saha Institute of Nuclear Physics, Homi Bhabha National Institute, Kolkata, India
\item \Idef{org111} School of Physics and Astronomy, University of Birmingham, Birmingham, United Kingdom
\item \Idef{org112} Secci\'{o}n F\'{\i}sica, Departamento de Ciencias, Pontificia Universidad Cat\'{o}lica del Per\'{u}, Lima, Peru
\item \Idef{org113} St. Petersburg State University, St. Petersburg, Russia
\item \Idef{org114} Stefan Meyer Institut f\"{u}r Subatomare Physik (SMI), Vienna, Austria
\item \Idef{org115} SUBATECH, IMT Atlantique, Universit\'{e} de Nantes, CNRS-IN2P3, Nantes, France
\item \Idef{org116} Suranaree University of Technology, Nakhon Ratchasima, Thailand
\item \Idef{org117} Technical University of Ko\v{s}ice, Ko\v{s}ice, Slovakia
\item \Idef{org118} The Henryk Niewodniczanski Institute of Nuclear Physics, Polish Academy of Sciences, Cracow, Poland
\item \Idef{org119} The University of Texas at Austin, Austin, Texas, United States
\item \Idef{org120} Universidad Aut\'{o}noma de Sinaloa, Culiac\'{a}n, Mexico
\item \Idef{org121} Universidade de S\~{a}o Paulo (USP), S\~{a}o Paulo, Brazil
\item \Idef{org122} Universidade Estadual de Campinas (UNICAMP), Campinas, Brazil
\item \Idef{org123} Universidade Federal do ABC, Santo Andre, Brazil
\item \Idef{org124} University of Cape Town, Cape Town, South Africa
\item \Idef{org125} University of Houston, Houston, Texas, United States
\item \Idef{org126} University of Jyv\"{a}skyl\"{a}, Jyv\"{a}skyl\"{a}, Finland
\item \Idef{org127} University of Kansas, Lawrence, Kansas, United States
\item \Idef{org128} University of Liverpool, Liverpool, United Kingdom
\item \Idef{org129} University of Science and Technology of China, Hefei, China
\item \Idef{org130} University of South-Eastern Norway, Tonsberg, Norway
\item \Idef{org131} University of Tennessee, Knoxville, Tennessee, United States
\item \Idef{org132} University of the Witwatersrand, Johannesburg, South Africa
\item \Idef{org133} University of Tokyo, Tokyo, Japan
\item \Idef{org134} University of Tsukuba, Tsukuba, Japan
\item \Idef{org135} University Politehnica of Bucharest, Bucharest, Romania
\item \Idef{org136} Universit\'{e} Clermont Auvergne, CNRS/IN2P3, LPC, Clermont-Ferrand, France
\item \Idef{org137} Universit\'{e} de Lyon, CNRS/IN2P3, Institut de Physique des 2 Infinis de Lyon, Lyon, France
\item \Idef{org138} Universit\'{e} de Strasbourg, CNRS, IPHC UMR 7178, F-67000 Strasbourg, France, Strasbourg, France
\item \Idef{org139} Universit\'{e} Paris-Saclay Centre d'Etudes de Saclay (CEA), IRFU, D\'{e}partment de Physique Nucl\'{e}aire (DPhN), Saclay, France
\item \Idef{org140} Universit\`{a} degli Studi di Foggia, Foggia, Italy
\item \Idef{org141} Universit\`{a} di Brescia, Brescia, Italy
\item \Idef{org142} Variable Energy Cyclotron Centre, Homi Bhabha National Institute, Kolkata, India
\item \Idef{org143} Warsaw University of Technology, Warsaw, Poland
\item \Idef{org144} Wayne State University, Detroit, Michigan, United States
\item \Idef{org145} Westf\"{a}lische Wilhelms-Universit\"{a}t M\"{u}nster, Institut f\"{u}r Kernphysik, M\"{u}nster, Germany
\item \Idef{org146} Wigner Research Centre for Physics, Budapest, Hungary
\item \Idef{org147} Yale University, New Haven, Connecticut, United States
\item \Idef{org148} Yonsei University, Seoul, Republic of Korea
\item \Idef{org150} Yukawa Institute for Theoretical Physics, Kyoto University, Kyoto, Japan
\end{Authlist}
\endgroup

\end{document}